\documentclass[amsmath,amssymb,superscriptaddress,prb,twocolumn]{revtex4-2}

\usepackage{graphicx}
\usepackage{multirow}
\usepackage{pdfpages}
\usepackage{hyperref}

\hypersetup{colorlinks=true,citecolor=blue,linkcolor=blue,urlcolor=blue}

\newcommand{\cm}{cm$^{-1}$}
\newcommand{\Ag}{$A_{\mathrm{g}}$}
\newcommand{\Eg}{$E_{\mathrm{g}}$}
\newcommand{\TaSe}{1$T$-TaSe$_2$}
\newcommand{\TCCDW}{$T_{\mathrm{CCDW}}$}
\newcommand{\TICCDW}{$T_{\mathrm{ICCDW}}$}

\makeatletter
\AtBeginDocument{\let\LS@rot\@undefined}
\makeatother

\begin{document}


\title{Identification of soft modes across the commensurate-to-incommensurate charge density wave transition in 1$T$-TaSe$_2$}


\author{M. Ruggeri}
    \affiliation{Università di Messina, Viale F. Stagno d'Alcontres 31, S. Agata, 98166 Messina, Italy}

\author{D. Wolverson}
    \affiliation{Centre for Nanoscience and Nanotechnology, Department of Physics, University of Bath, Bath BA2 7AY, United Kingdom}

\author{V. Romano}
    \affiliation{Dipartimento di Fisica, Politecnico di Milano, Milan 20133, Italy}

\author{G. Cerullo}
    \affiliation{Dipartimento di Fisica, Politecnico di Milano, Milan 20133, Italy}

\author{C. J. Sayers}
    \email[Correspondence email address: ]{charles.sayers@polimi.it}
    \affiliation{Dipartimento di Fisica, Politecnico di Milano, Milan 20133, Italy}

\author{G. D'Angelo}
    \email[Correspondence email address: ]{giovanna.dangelo@unime.it}
    \affiliation{Università di Messina, Viale F. Stagno d'Alcontres 31, S. Agata, 98166 Messina, Italy}


\begin{abstract}

\TaSe~is a prototypical charge density wave (CDW) material for which electron-phonon coupling and associated lattice distortion play an important role in driving and stabilizing the CDW phase. Here, we investigate the lattice dynamics of bulk \TaSe~using angle-resolved ultralow wavenumber Raman spectroscopy down to 10 \cm. Our high-resolution spectra allow us to identify at least 27 Raman-active modes in the commensurate (CCDW) phase. Contrary to other layered materials, we do not find evidence of interlayer breathing or shear modes, suggestive of $AA$ stacking in the bulk, or sufficiently weak interlayer coupling. Polarization dependence of the mode intensities allows the assignment of their symmetry, which is supported by first-principles calculations of the phonons for the bulk structure using density functional theory. A detailed temperature dependence in the range $T$ = 80 - 500 K allows us to identify soft modes associated with the CDW superlattice. Upon entering the incommensurate (ICCDW) phase above 473 K, we observe a dramatic loss of resolution of all modes, and significant linewidth broadening associated with a reduced phonon lifetime as the charge-order becomes incommensurate with the lattice.

\end{abstract}


\maketitle


\section{Introduction}

Understanding the fundamental origin of strongly correlated phenomena and their interplay in complex systems is an important goal of condensed matter physics. The layered transition metal dichalcogenides (TMDs) based on tantalum, e.g. TaS$_2$ and TaSe$_2$, are an ideal platform on which to study such phenomena owing to their extremely rich phase diagrams comprised of multiple charge density wave (CDW) transitions, Mott insulating phases, and superconductivity \cite{Wilson1975,Sipos2008}. Control of these various phases using external parameters has already been demonstrated via doping \cite{Liu2016}, and pressure \cite{Sipos2008,Lin2021}, in addition to great tunability with sample thickness beyond the monolayer limit towards tens of nanometres \cite{Yoshida2014}.

A key concept in CDW physics is that of commensurability, whereby several energy-minimizing symmetry states can be realised in the same system, as the spatially modulated charge density can assume different periodicities that are either commensurate or incommensurate with the underlying lattice. Incommensurability has been linked, for example, to the emergence of superconductivity at domain walls \cite{Kogar2017,Yan2017}. While in the Ta-based TMDs, incommensurate-to-commensurate transitions lead to changes in the electrical resistance by several orders of magnitude with significant hysteresis \cite{Wilson1975}. Understanding this behaviour is important for realising optoelectronics applications based on CDW devices such as memristors \cite{Yoshida2015,Vaskivskyi2016,Mraz2022}.

In general, the group-5 TMDs, i.e. those with transition metal V, Nb, or Ta, are believed to exhibit CDWs which are predominantly related to a structural phase transition driven by electron-phonon coupling \cite{Rossnagel2011,Zhu2015}. Indeed, it has recently been demonstrated that the lattice degrees of freedom play a particularly important role in \TaSe, as it can fully explain its ground state electronic properties, including the opening of the band gap, and stabilizing the CDW phase by a minimization of the lattice energy \cite{Sayers2023}. By studying the crystal lattice using traditional probes, such as diffraction \cite{Wilson1975} or Raman spectroscopy \cite{Xi2015}, or by directly observing coherent phonon oscillations in the time domain \cite{Perfetti2006,Sayers2020,Sayers2022}, it is possible to search for signatures of new modes that arise as a result of the lattice reconstruction or soft mode behaviour near the critical temperature.

\begin{figure}[ht]
    \includegraphics[width=1.0\linewidth]{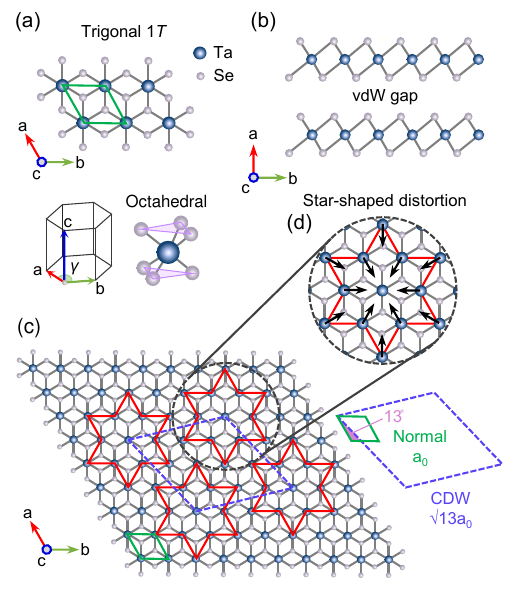}
    \caption{\label{fig:structure} \textbf{Crystal structure of \TaSe~and the CDW lattice distortion.} (a) View of the layer plane showing the primitive unit cell of undisorted trigonal (1$T$) TaSe$_2$ and its octahedral coordination. (b) Side view showing the three atomic planes in the configuration Se-Ta-Se which comprise each layer, separated by the van der Waals (vdW) gap, with $AA$ stacking. (c) View of the layer plane showing a comparison of the unit cell of the normal phase (green) and the reconstructed $(\sqrt{13}a_0 \times \sqrt{13}a_0)R_{\mathrm{13^{\circ}}}$ CDW phase (blue). (d) Illustration of the displacement of the 12 neighbouring atoms surrounding a central Ta atom, forming a star-shaped configuration.}
\end{figure}

In this work, we investigate the lattice dynamics of bulk \TaSe~using ultralow wavenumber Raman spectroscopy down to 10 \cm. Our high-resolution data reveals an extremely rich Raman spectrum in the commensurate CDW phase. Particular attention has been paid to the low-frequency region below 120 \cm~where the spectrum contains Raman-active phonons of the reconstructed lattice and CDW collective excitations \cite{Sugai1981}, as well as the expected interlayer breathing (LBM) and shear (LSM) modes of TMDs \cite{Yan2015}. In addition, we report on the frequency region between 140 \cm~and 300 \cm~where the vibrations of the normal structure are expected \cite{Samnakay2015}. By performing a polarization dependence of the mode intensities, and combining our results with density functional theory (DFT) calculations of the phonon frequencies of bulk \TaSe, we identify the vibrational symmetry of each mode. Then, by performing a detailed temperature dependence in the range $T$ = 80 - 500 K, we identify the soft modes belonging to the CDW superlattice and investigate their behaviour across the incommensurate-to-commensurate CDW transition.

The crystal structure of undistorted trigonal (1$T$) TaSe$_2$ is shown in Figures~\ref{fig:structure}(a) and~\ref{fig:structure}(b). At room temperature, however, \TaSe~exists in the commensurate (CCDW) phase in which the crystal lattice is distorted to form a star-shaped configuration whereby the 12 neighbouring Ta atoms at the tips of the star move towards the atom at its centre, as illustrated in Figure~\ref{fig:structure}(c) and Figure~\ref{fig:structure}(d). The distortion results in a new superlattice periodicity defined by a unit cell of $\sqrt{13}a_0 \times \sqrt{13}a_0$ and rotated by 13° ($R_{\mathrm{13^{\circ}}}$). When heated above \TCCDW~= 473 K, it undergoes a first-order phase transition from the CCDW to an incommensurate (ICCDW) state as the charge-order becomes incommensurate with the underlying lattice \cite{DiSalvo1974}. The transition from the ICCDW to normal state is then predicted at higher temperatures, \TICCDW~= 600 K, although the 1$T$ structure first becomes unstable above 530 K, where it undergoes an irreversible interpolytypic transformation from 1$T$ to 3$R$ \cite{Huisman1969,Wilson1975}, making the normal phase of \TaSe~experimentally inaccessible. The new periodicity dictated by the CDW unit cell results in the folding of phonon branches across the Brillouin zone (BZ) to the $\Gamma$-point ($k = 0)$, and hence several new Raman-active modes arise in the reconstructed phase \cite{Wilson1975,Smith1976,Tsang1977,Sugai1981,Samnakay2015}.


\section{Results \& Discussion}

In its undistorted structure for $T >$ \TICCDW, bulk \TaSe~belongs to the $D_{\mathrm{3d}}$ point group \cite{Samnakay2015}, for which the irreducible representation gives just two discrete Raman-active modes; $A_{\mathrm{1g}}$ and \Eg. These normal phase modes are expected to appear in the high-frequency region 140 - 300 \cm~\cite{Tsang1977}. In the distorted CDW phase structure for $T <$ \TICCDW, it assumes the triclinic $C_{i}$ point group \cite{Samnakay2015} with a dramatic increase in the number of optical modes due to BZ folding up to a total of 57 Raman-active \Ag-like modes. In principle, there should be no \Eg~modes in the $C_{i}$ point group, however earlier experimental work showed that the mode symmetries in \TaSe~can be readily classified in terms of both \Ag~and \Eg, which is valid if the interlayer coupling is weak enough such that the trigonal $C_{3i}$ point group determines the Raman tensors \cite{Sugai1981}. As will be discussed later, we also observe both \Ag~and \Eg~modes in the Raman spectrum, hence we assume the $C_{3i}$ point group for the CCDW structure which should consist of a total of 57 Raman-active modes, or 38 discrete frequencies; 19 \Ag- and 19 \Eg-symmetry, where the latter are always doubly-degenerate.

\begin{figure*}[t]
    \includegraphics[width=1.0\linewidth]{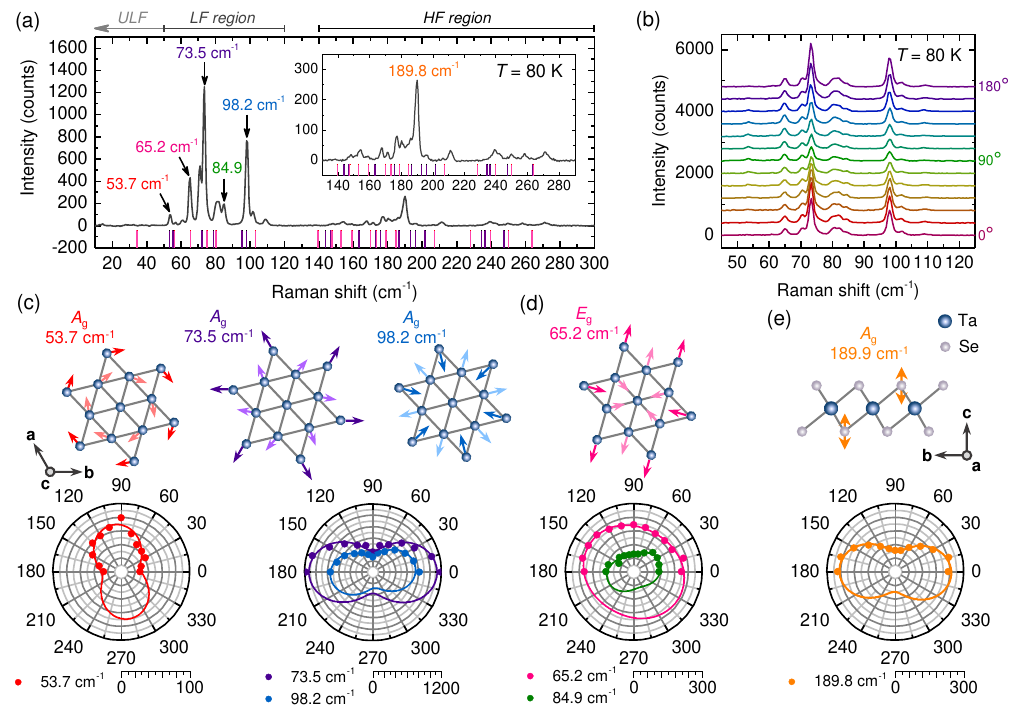}
    \caption{\label{fig:assignment} \textbf{Identification of Raman-active modes in the CCDW phase of \TaSe.} (a) High-resolution Raman spectrum measured at $T$ = 80 K ($\theta$ = 0°) where a total of 27 well-resolved peaks can be identified. Particular modes of interest are labelled. The horizontal arrows on the top axis indicate the high (HF), low (LF) and ultra-low frequency (ULF) regions. The inset shows a zoom of the HF region. The vertical bars (arbitrary height) represent predicted phonon frequencies from first-principles calculations with \Ag~(purple) or \Eg~(pink) symmetry. (b) Angle-resolved Raman spectra at  $T$ = 80 K for excitation polarization angle $\theta$ = 0° - 180°. The spectra are offset for clarity. (c) Identification of selected LF modes with \Ag~symmetry based on polar plots of the mode intensities as a function of polarization angle. The radial axis is the Raman intensity (counts) represented by the horizontal scale bar. Sketches of the corresponding atomic displacements of Ta atoms in the star cluster obtained by DFT are shown for each mode. Panels (d) and (e) show polar plots and displacement sketches corresponding to selected LF \Eg~modes and HF \Ag~modes, respectively.}
\end{figure*}

We start by analysing the high-resolution Raman spectrum of \TaSe~in the CCDW phase at 80 K for parallel polarization, $\theta$ = 0° (where $\theta$ is the angle between the incident and the detected light) over the frequency range 10 - 300 \cm, as shown in Figure~\ref{fig:assignment}(a). The spectrum can be separated into three distinct regions; (i) the ultralow frequency (ULF) region below 50 \cm~where the interlayer LBM and LSM modes are expected, (ii) the low frequency (LF) region 50 - 120 \cm~in which the modes are understood to be associated with the CDW phase, i.e. vibrations of the distorted lattice originating from the folding of the BZ and collective CDW excitations, (iii) the high frequency (HF) region 140 - 300 \cm~in which the modes arise from out-of-plane (\Ag) or in-plane (\Eg) intralayer vibrations of Se atoms around each Ta atom \cite{Sugai1981}. From our experimental data in Figure~\ref{fig:assignment}(a), we are able to resolve at least 27 modes out of the 38 predicted modes with discrete frequencies, presumably because some are intrinsically weak making them undetectable based on our instrument sensitivity, while others may be overlapped in frequency if they are nearly-degenerate. 

To assist in identifying the observed modes, we performed first-principles calculations using DFT to obtain the phonon frequencies and their corresponding symmetries, as described in the Methods section. We computed the phonon dispersion of bulk \TaSe~with $AA$ stacking, including van der Waals interactions between the layers. As the input, we used the trigonal CCDW phase structure ($C_{3i}$ point group), which we found to be stable as evidenced by the absence of imaginary phonon frequencies across the entire three-dimensional BZ (see Supplemental Material A \cite{Supplemental}). The predicted phonon frequencies are shown as vertical bars in Figure~\ref{fig:assignment}(a) and reported in full in Supplemental Material B \cite{Supplemental}. To gain additional information about the mode symmetries, and to confirm the predictions of DFT, we performed angle-resolved Raman measurements at $T$ = 80 K by rotating the polarization, $\theta$ of the scattered light from 0 to 180°, as shown in Figure~\ref{fig:assignment}(b). In backscattering normal to the layer, \Ag~modes exhibit a two-fold symmetry with intensity maxima for parallel polarization configuration ($\theta$ = 0° and $\theta$ = 180°), whereas they are attenuated for cross-polarization ($\theta$ = 90°) \cite{Sugai1981,Hart2016,Lacinska2022}. On the other hand, \Eg~modes can be detected in both co- and cross-polarization, and hence exhibit a fully symmetric circular shape. Figures~\ref{fig:assignment}(c) - (e) report the polarization angle-dependent intensities as polar plots, where the solid line is a fit to the data using $I \propto \cos^{2}(\theta)$ \cite{Hart2016,Lacinska2022}. 

Returning to the LF region of Figure~\ref{fig:assignment}(a), we observe 12 resolvable peaks, and highlight several modes of interest. Based on their two-fold symmetry (see Figure~\ref{fig:assignment}(c)), we assign the modes at 53.7, 73.5, and 98.2 \cm~as \Ag. While, based on their circular symmetry (see Figure~\ref{fig:assignment}(d)), the modes at 65.2 and 84.9 \cm~are \Eg. The presence of these modes, and others which exhibit significant \Eg-like character (see Supplemental Material B \cite{Supplemental}), suggests $C_{3i}$ structure in bulk \TaSe~\cite{Sugai1981}. The most intense LF modes, found at 73.5 and 98.2 \cm, are believed to be modes of the CDW. Their large Raman intensity, which is $\sim$ 3 - 5 times greater than the most intense HF modes, can be considered an indication of the exceptionally strong electron-phonon coupling (EPC) associated with these modes, particularly when compared to other TMDs with 1$T$ symmetry, such as TiSe$_2$ or VSe$_2$, in which the opposite trend is found, i.e. the CDW modes are typically less intense and broader than the normal state modes \cite{Holy1977,Sugai1985}. To further assess EPC, we analyzed the absolute magnitude of perturbations to the electronic band structure induced by atomic displacements along the eigenvectors of the CCDW $\Gamma$-point phonons based on the first-principles calculations (see Supplemental Material C \cite{Supplemental}). We find a good agreement between the most intense Raman-active modes and the magnitude of the induced perturbation of the Fermi energy, particularly for the modes at 73.5 and 98.2 \cm, indicating a strong EPC. This assessment is consistent with time-domain experiments whereby these modes dominate the transient optical response of \TaSe~\cite{Sayers2022} and directly modulate the valence band energy (bandgap) due to a charge migration between atoms at the tips and the centre of the star pattern \cite{Sayers2023}.

Of particular importance is the mode found at 73.5 \cm~which, based on its frequency, we identify as the breathing mode of the star-shaped distortion \cite{Perfetti2006,Albertini2016,Sayers2020,Sayers2023}. The atomic displacement associated with this mode frequency, as obtained from DFT calculations, confirms this to be the case as there is a collective in-plane expansion of all 12 neighbouring Ta atoms around the central Ta atom, with an additional rotational component as the inner and outer hexagons of the star cluster twist slightly in opposite directions during the expansion, as illustrated in Figure~\ref{fig:assignment}(c). Similarly, we find that the intense 98.2 \cm~mode also corresponds to a type of breathing mode, but with the inner atoms of the star expanding outwards while the outer atoms contract towards the centre, with no sizeable rotation. Finally, we identify the lowest frequency mode at 53.7 \cm~with weaker star expansion but with a significant rotational component whereby all 12 atoms twist in the same direction around the centre, giving rise to a well-defined chirality. The predicted atomic displacement confirms this mode to be an \textit{intralayer} vibration, and hence we rule out that this mode is either the LBM or LSM mode shifted to high frequencies. In contrast to what was reported by Sugai \textit{et al.} \cite{Sugai1981}, we find that the mode at 53.7 \cm~has \Ag~symmetry, while the mode at 60.8 \cm~is \Eg. We note that both of these modes have been observed in the time-domain, and their \Ag~and \Eg~symmetry assignments here are in excellent agreement with the reported displacive and impulsive excitation processes, respectively, as determined by phase analysis \cite{Sayers2022,Melnikov2011}.

In the HF region of Figure~\ref{fig:assignment}(a), we observe 15 resolvable peaks and highlight the most intense at 189.8 \cm, which we believe to be the \Ag~mode of the normal structure \cite{Tsang1977}, which corresponds to the out-of-plane vibration of Se atoms around the central Ta atomic plane, as illustrated in Figure~\ref{fig:assignment}(e).

Finally, we inspect the ULF region of Figure~\ref{fig:assignment}(a) for signatures of interlayer breathing (LBM) and shear (LSM) modes, which are typical features of van der Waals compounds such as graphene \cite{Lui2012,Tan2012}, phosphorene \cite{Ling2015,Mao2021}, and TMDs \cite{Zhao2013,He2016} found below 50 \cm. Contrary to what is found in its 2$H$ phase \cite{Lin2020}, we detected no signatures of the LBM and LSM for bulk \TaSe~down to 10 \cm. The absence of interlayer modes may suggest $AA$ stacking in the bulk, similar to what was predicted by Yan \textit{et al.} \cite{Yan2015}. In this scenario, the motion of the layers is effectively in-phase as a result of the single layer per unit cell; hence, the LBM and LSM tend to zero frequency at the $\Gamma$-point. Alternatively, if there is sufficiently weak interlayer coupling \cite{Sugai1981}, the interlayer modes will shift to low frequency, making them undetectable here if $\omega <$ 10 \cm~and hence defines the upper limit. In the case of very weak interlayer coupling, bulk \TaSe~effectively behaves like an isolated monolayer. As mentioned previously, the presence of \Eg~modes in the Raman spectrum suggests $C_{3i}$ symmetry in the bulk. Because $C_{3i}$ is the point group of a monolayer, and $AA$ stacking corresponds to monolayers aligned vertically with identical star positions (centre-on-centre) \cite{Wang2023}, the symmetries of a monolayer-like object or bulk $AA$ are the same. Hence, one can imagine similar implications for the electronic properties of \TaSe~for either the $AA$ stacking or weak interlayer coupling scenarios. We note that layer stacking order, layer dimerization, and dimensionality effects are believed to have broad implications for understanding the electronic structure of of the Ta-based TMDs including; the origin of bandgaps in the ground state, a debate on bulk metallicity, and the establishment of insulating states previously linked to a CCDW-Mott phase \cite{Wang2023,Ritschel2018,Lee2019,Butler2020,Chen2020,Fei2022,Ramos2023,Tian2024}.

\begin{figure*}[t]
    \includegraphics[width=1.0\linewidth]{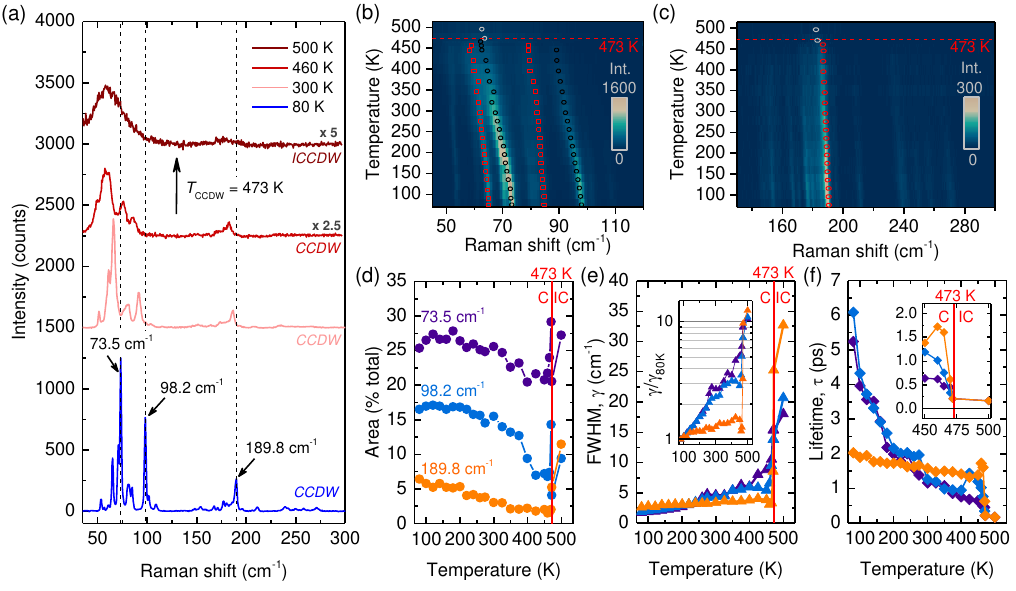}
    \caption{\label{fig:temperature} \textbf{Temperature-dependence of the Raman-active modes across the CCDW-to-ICCDW transition in \TaSe.} (a) Raman spectra at selected temperatures. Spectra are offset for clarity. The vertical dashed lines indicate the position of the 73.5 \cm, 98.2 \cm, and 189.8 \cm~peaks corresponding to their frequency at 80 K. The intensity of the spectra at 460 K and 500 K have been multiplied by a factor of 2.5 and 5, respectively. (b) Contour plot for the LF region, and (c) HF region, showing the temperature dependence of the Raman spectra over the range $T$ = 80 - 500 K. A Voigt fitting procedure is used to evaluate the spectra at each temperature (see Supplemental Material B \cite{Supplemental}). Black (below \TCCDW = 473 K) and grey (above \TCCDW) circles highlight the frequencies of the strongly temperature-dependent 73.5 \cm~and 98.2 \cm~modes, while the red markers indicate the 65.2 \cm, 84.9 \cm~and 189.8 \cm~modes. (d) Temperature dependence of the peak areas, evaluated as a percentage of the total area of all observed peaks. The red vertical line indicates the transition temperature (473 K) between the incommensurate (IC) and commensurate (C) CDW phases (e) Temperature dependence of the FWHM and (f) corresponding phonon lifetime $\tau = \hbar / \gamma_L$ of selected modes based on the Lorentzian linewidth, $\gamma_L$.}
\end{figure*}

Now that we have identified the Raman-active phonon modes of \TaSe~in the CCDW phase and their associated symmetries, we next investigate their temperature dependence across the CCDW-to-ICCDW phase transition. Figure~\ref{fig:temperature}(a) shows the Raman spectra for selected temperatures above and below the expected phase transition temperature at \TCCDW~= 473 K. Starting from 80 K, we observe a significant redshifting of all modes with increasing temperatures, while those in the LF region appear to shift more than those in the HF region. We also observe a reduction in the maximum Raman intensity and significant linewidth broadening. At 460 K, i.e. for $T<$ \TCCDW, several phonon modes can still be distinguished in the LF region. However, as the temperature is increased further to 500 K, i.e. for $T >$\TCCDW, they are no longer well-resolved and combine to form a broad continuum centred around $\sim$ 62 \cm. This sudden broadening is evidence of the first-order CCDW-to-ICCDW phase transition \cite{Tsang1977,Sugai1981,Wang2020,Sayers2020}. Such behaviour has also been observed in 1$T$-TaS$_2$ following the loss of charge-lattice commensurability when passing from the CCDW phase to the nearly-commensurate (NCCDW) phase \cite{Smith1976}.

The full temperature dependence of the Raman spectra in the range $T$ = 80 - 500 K are shown as contour plots for the LF and HF regions in Figures~\ref{fig:temperature}(b) and (c), respectively. To track the evolution of the spectral features with temperature, we performed a Voigt lineshape fitting procedure (see Supplemental Material B \cite{Supplemental}). The peak frequencies of selected modes obtained from the Voigt fitting are shown as markers in Figures~\ref{fig:temperature}(b) and (c). As mentioned previously, all peaks redshift with increasing temperature. However, in the LF region, the modes at 73.5 \cm~and 98.2 \cm~exhibit a strong temperature dependence, particularly when approaching \TCCDW, where the change in frequency of the 73.5 \cm~star breathing mode is as large as $\Delta \omega \approx$ 14 \cm~between 80 K and 460 K. Other modes, such as those labelled 65.2 \cm~and 84.9 \cm, are more weakly temperature dependent. As a reference, we use the 189.8 \cm~\Ag~mode of the normal structure found in the HF region, which exhibits a shift of only $\Delta \omega \approx$ 7 \cm~over the same temperature range. Based on this, we identify the 73.5 \cm~and 98.2 \cm~as \textit{soft modes} of the CDW. This is fully consistent with our DFT calculations of the atomic motion associated with these mode frequencies, which predict an expansion-contraction of the star pattern in both cases, equivalent to the lattice displacement between CDW and normal structures, as illustrated in Figure~\ref{fig:assignment}(c).

To give a further quantitative description of the behaviour of the Raman modes involved in the CDW phase transition, we analyzed the temperature dependence of the relevant Voigt lineshape parameters, including peak area and width. In general, as shown in Figure~\ref{fig:temperature}(d) and (e), we see a reduction in the peak areas of all modes with increasing temperature, while the full-width-half-maximum (FWHM) shows the opposite trend. Figure~\ref{fig:temperature}(d) shows that the 189.8 \cm~normal phase mode exhibits a linear reduction in the peak area from 80 to 470 K, while the 73.5 \cm~and 98.2 \cm~CDW modes are clearly non-linear with temperature, which in the absence of the CCDW-to-ICCDW transition, would be expected to fall to zero at the normal state transition at 600 K, as observed in 2$H$-TaSe$_2$ \cite{Sugai1981}. This non-linear behaviour can be explained by thermally activated screening of the electron-phonon interaction, which reduces the CDW mode intensity.

Figure~\ref{fig:temperature}(e) shows the temperature dependence of the peak widths (FWHM). Initially at 80 K, the CDW modes are narrower than the normal state modes (indicating their longer lifetime) with a FWHM of $\gamma_\mathrm{80K} \approx$ 1.7 \cm~and 1.9 \cm, for the 73.5 \cm~and 98.2 \cm~modes, respectively, while it is $\approx$ 2.6 \cm~for the 189.9 \cm~mode. With increasing temperature, the CDW modes exhibit a non-linear increase in the linewidth and become broader than the normal state modes above $\sim$250 K. This is more clearly illustrated in the inset of Figure~\ref{fig:temperature}(e), where the linewidth is normalised to the 80 K value and shows that between 80 and 450 K, the 73.5 \cm~and 98.2 \cm~modes exhibit a broadening of $\sim$5 and 3 times, respectively, while the linewidth of the 189.9 \cm~mode increases by only $\sim$1.5 times.

In the ICCDW phase, as mentioned previously, all modes broaden significantly and merge to form two broad features in the vicinity of the original LF and HF modes due to the sudden breaking of translational symmetry \cite{Samnakay2015}. Indeed, deconvolution by the Voigt fitting analysis reveals that the 73.5 \cm, 98.2 \cm~and 189.8 \cm~modes reach $\geq 10$ times their width at 80 K. The Raman linewidth provides an indication of the intrinsic phonon lifetime by considering $\tau = \hbar / \gamma_L$, where $\gamma_L$ is the Lorentzian linewidth \cite{Bergman1999}. In Figure~\ref{fig:temperature}(f), the lifetime of the 73.5 \cm~CDW mode at 80 K was found to be $\tau$ = 5.3 ps, which is in excellent agreement with that measured directly in the time-domain by ultrafast spectroscopy \cite{Perfetti2006,Sayers2020,Sayers2022}. Between 80 and 450 K, the normal phase 189.8 \cm~mode shows a relatively constant lifetime of 1.5 - 2.0 ps, whereas the 73.5 \cm~CDW mode drops dramatically by an order of magnitude from $\sim$ 5 to 0.5 ps. Close to the CCDW-to-ICCDW transition, as shown in the inset of Figure~\ref{fig:temperature}(f), the lifetime of all modes falls sharply towards a few hundreds of femtoseconds, which is comparable to, or less than, one oscillatory period of the 73.5 \cm~($1/\omega$ = 450 fs) and 189.8 \cm~($1/\omega$ = 175 fs) modes, indicative of a crossover to critical damping at \TCCDW.


\section{Conclusion}

Using a combination of high-resolution Raman spectroscopy measurements and first-principles calculations, we have identified the soft modes associated with the CDW in bulk \TaSe. Deep in the commensurate phase, these CDW modes exhibit exceptionally large Raman intensity and narrow linewidths, indicative of their strong electron-phonon coupling and long lifetimes. The calculated atomic displacements of these modes confirm them to be predominantly associated with an expansion-contraction motion of the star-shaped lattice distortion. We find no evidence of the interlayer breathing or shear modes down to 10 \cm, suggestive of $AA$ stacking in the bulk or weak interlayer coupling, which has implications for understanding electron correlations in Ta-based TMDs where the stacking order, layer dimerization, dimensionality effects have been suggested to be fundamentally important in establishing the low temperature CCDW-Mott phase.


\section*{Methods}

\subsection{Raman spectroscopy}

Single crystals of \TaSe~were grown using chemical vapour transport according to the methods in Ref.~\cite{Sayers2020}. Micro-Raman spectroscopy measurements were performed using a HORIBA LabRam HR 800 spectrometer in backscattering geometry, equipped with a low-wavenumber filter allowing acquisition of spectra down to $\sim$ 10 \cm. A liquid nitrogen flow cryostat was used to control the sample temperature in the range $T$ = 80 – 500 K. Prior to measurements, the \TaSe~sample was cleaved to expose a pristine surface. A 532 nm laser was used for excitation, which was focused onto the sample surface using a 50x objective lens. The measured laser power on the sample was 32 $\mu$W. The scattered light was dispersed by a 1800 gr/mm grating and detected by a CCD camera cooled with liquid nitrogen. The spectral resolution was $\simeq$ 0.5 \cm. Angle-resolved Raman spectra were acquired by keeping the incident light polarization fixed and selecting the polarization of the collected scattered light using a waveplate in the range $\theta$ = 0 to 180°, with a step of 15°. To account for possible intensity variations due to polarization-sensitive throughput of the detection optics, a baseline subtraction was performed to all data.

\subsection{First-principles calculations}

Phonon frequencies and displacement patterns were calculated using density functional theory (DFT) with a plane-wave basis set and PBE-GGA \cite{Perdew1996} projector-augmented wave pseudopotentials, as implemented in VASP \cite{Kresse1999,Hafner2008}. The frozen phonon approach of Phonopy \cite{Togo2015} was used to evaluate force constants over a single unit cell of the trigonal CCDW phase structure ($C_{3i}$ point group) with $AA$ stacking comprising 39 atoms, with lattice parameters $a$ = $b$ = 12.607~\AA~and $c$ = 6.244~\AA. A supercell was not used since the CCDW phase structure is already a supercell of the normal phase. A kinetic energy cutoff of 300 eV and Monkhorst-Pack $k$-point grid of 6$\times$6$\times$10 \cite{Monkhorst1976} gave well-converged results and real phonon frequencies over the entire three-dimensional Brillouin zone. Van der Waals interactions between layers were modelled using the Grimme-D3 approach \cite{Grimme2010}. Similar results (matching displacement patterns and phonon frequencies consistent to within expected accuracies of $\sim$2~cm$^{-1}$) were obtained using density functional perturbation theory, as implemented in Quantum Espresso \cite{Baroni2001}. Results obtained using PZ-LDA \cite{Purdew1981} without van der Waals corrections and using ultrasoft pseudopotentials \cite{Vanderbilt1990} shows that there is some re-ordering of modes that lie close in frequency, but that similar displacement patterns are obtained for the normal modes, which allows comparison of results obtained with different pseudopotentials. The overall frequency range of the normal modes at the $\Gamma$-point is comparable for all choices of pseudopotential.


\section{Acknowledgements}

Computational work was performed on the University of Bath’s High Performance Computing Facility and was supported by the EU Horizon 2020 OCRE/GEANT project ``Cloud funding for research”. Data associated with this study are openly available from the University of Bath data archive at the following DOI: \href{http://doi.org/10.15125/BATH-01357}{10.15125/BATH-01357}


\bibliographystyle{unsrt}

\newpage


\section*{Supplementary material}

# Supplemental Material for: "Identification of soft modes across the commensurate-to-incommensurate charge density wave transition in $1T$-TaSe$_2$"

M. Ruggeri,[1] D. Wolverson,[2] V. Romano,[3] G. Cerullo,[3] C. J. Sayers,[3, *] and G. D'Angelo[1, †]

[1]*Università di Messina, Viale F. Stagno d'Alcontres 31, S. Agata, 98166 Messina, Italy*

[2]*Centre for Nanoscience and Nanotechnology, Department of Physics,*
*University of Bath, Bath BA2 7AY, United Kingdom*

[3]*Dipartimento di Fisica, Politecnico di Milano, Milan 20133, Italy*

_____

* Correspondence email address: charles.sayers@polimi.it

† Correspondence email address: giovanna.dangelo@unime.it

## A. PHONON DISPERSION FROM FIRST-PRINCIPLES CALCULATIONS

Figure S1 shows the phonon dispersion of bulk $1T$-TaSe$_2$ obtained from first-principles calculations using density functional theory (DFT) which includes van der Waals interactions between the layers. Full details are reported in the methods section of the main text. As the input to the calculations, we used the trigonal CCDW phase structure ($C_{3i}$ point group) with $AA$ stacking. Crucially, we find that this structure is stable, as evidenced by the absence of negative (imaginary) phonon frequencies across the entire three-dimensional Brillouin zone.

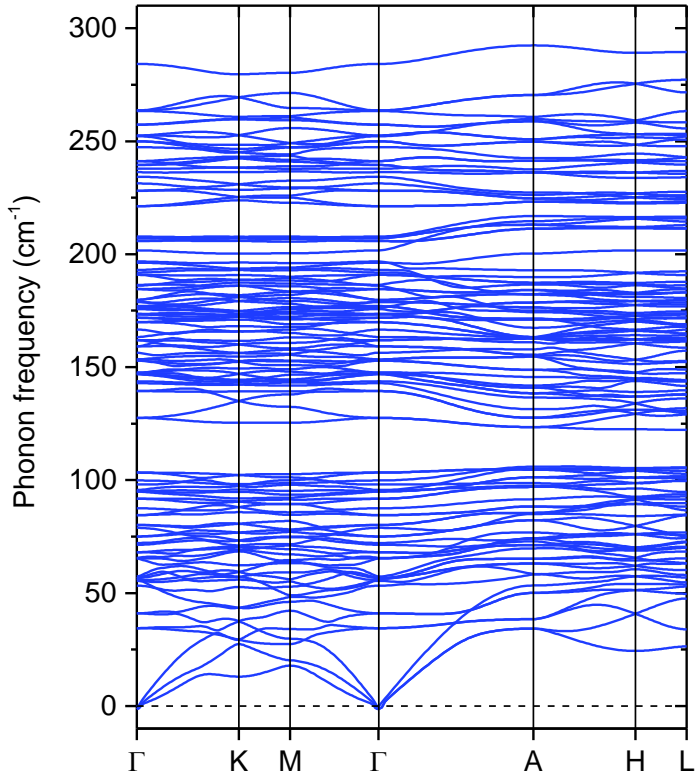

FIG. S1. **Phonon dispersion of bulk $1T$-TaSe$_2$ with $AA$ stacking.** Calculations were performed using the trigonal CCDW phase structure ($C_{3i}$ point group) including van der Waals interactions between the layers. The horizontal dashed line indicates zero frequency.

## B. ASSIGNMENT OF PHONON MODE SYMMETRIES

As described in the main text, we observe a total of 27 well-resolved modes in the Raman spectrum measured at 80 K. To obtain the frequencies, we fit the data with Voigt lineshapes as shown in Figure S2. Then to obtain the symmetries of these modes, we analyzed their intensity as a function of the excitation polarization. The results are reported in Tables S1 and S2 for the low (50 - 120 cm$^{-1}$) and high frequency (140 - 300 cm$^{-1}$) regions respectively, where the modes are numbered in order of ascending frequency (1st column). The centre frequencies from the Voigt fitting are given (2nd column) and their assigned symmetries (3rd column) based on the corresponding polar plots (4th column). Note that the mode at 56.9 cm$^{-1}$ is too weak to reliably assign its symmetry. Based on our computational approach, we find a total of 57 Raman-active modes; 19 $A_g$+ 19 $E_g$(2), where the Eg-type are all doubly-degenerate, hence giving 38 discrete frequencies. In the low frequency (LF) region (Figure S1), we predict 12 discrete Raman-active modes from DFT in good agreement with the 12 modes found experimentally, although these are not precisely matched. In the high frequency (HF) region (Figure S2), we predict 26 discrete Raman-active modes from DFT compared to 15 found experimentally. In Tables S1 and S2, we match the experimental and calculated modes when there is a reasonable agreement with both the predicted frequency (5th column) and symmetry (6th column). Finally, we show the corresponding atomic displacements, both in-plane i.e. *ab* plane (7th column) and out-of-plane i.e. *c*-axis (8th column), as predicted by DFT and visualized in *Jmol*. The atoms are shown in their equilibrium positions. The coloured arrows indicate the direction of motion, while their length is representative of the magnitude of displacement. For the LF modes, we only show Ta atoms (blue) of the star cluster for clarity. For the HF modes in the out-of-plane direction, we also show Se atoms (orange) comprising the entire Se-Ta-Se layer. We note that since all $E_g$ modes are doubly-degenerate, for simplicity in Table S1 each identical $E_g$ pair is represented by a single frequency and one of the two possible corresponding atomic vibrations is shown (the other being identical but with the opposite direction of motion).

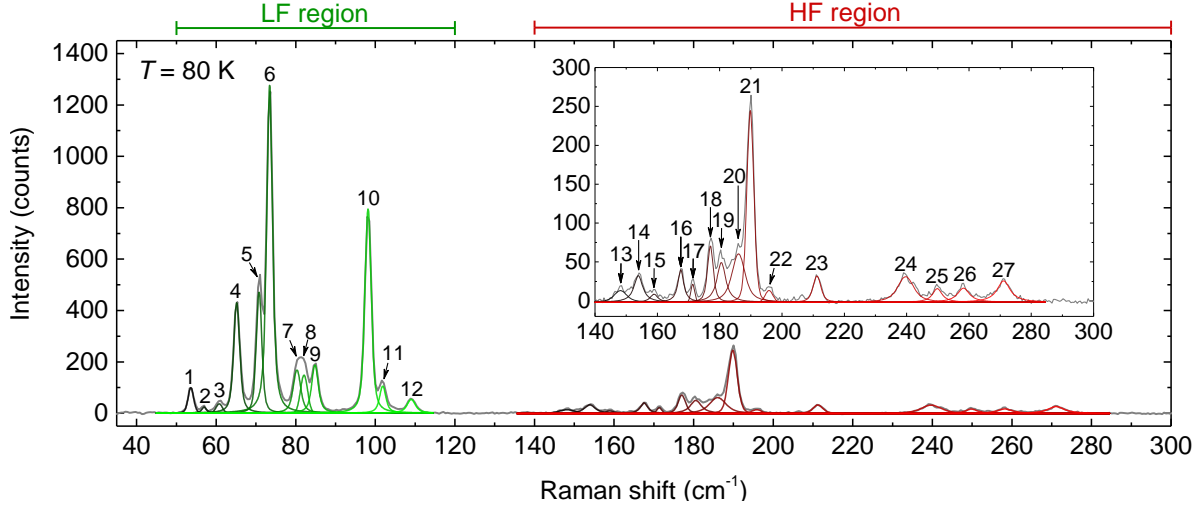

FIG. S2. **Voigt lineshape fitting procedure of Raman spectra.** The grey curve is the high resolution Raman spectrum of $1T$-TaSe$_2$ at $T = 80$ K ($\theta = 0°$). The inset shows a zoom of the high frequency region. The coloured curves correspond to the total of 27 peaks fitted with Voigt lineshapes; shades of green for the low frequency region (50 - 120 cm$^{-1}$) containing modes 1 - 12, and red for the high frequency region (140 - 300 cm$^{-1}$) containing modes 13 - 27.

TABLE S1: **Assignment of the low frequency modes (50 - 120 cm$^{-1}$).** Selected modes of interest (no. 1, 4, 6, 9 and 10) are highlighted in bold.

| Experiment | | | | Theory | | | |
|---|---|---|---|---|---|---|---|
| Mode | $\omega$ (cm$^{-1}$) | Symm. | Polar | $\omega$ (cm$^{-1}$) | Symm. | Disp. $ab$ | Disp. $c$ |
| – | – | – | – | 34.43 | $E_{\mathrm{g}}$ | | |
| – | – | – | – | 53.21 | $A_{\mathrm{g}}$ | | |
| **1** | **53.7** | $A_{\mathrm{g}}$ | | 55.40 | $A_{\mathrm{g}}$ | | |

| Experiment | | | | Theory | | | |
|---|---|---|---|---|---|---|---|
| Mode | $\omega$ (cm$^{-1}$) | Symm. | Polar | $\omega$ (cm$^{-1}$) | Symm. | Disp. $ab$ | Disp. $c$ |
| 2 | 56.9 | weak | n/a | 56.31 | $E_g$ | 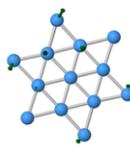 | 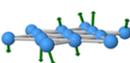 |
| 3 | 60.8 | $E_g$ | 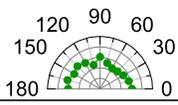 | – | – | – | – |
| **4** | **65.2** | $E_g$ | 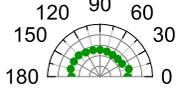 | 65.44 | $E_g$ | 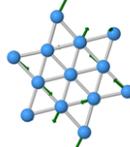 | 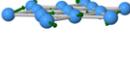 |
| 5 | 70.8 | $E_g$ | 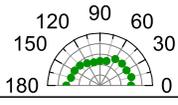 | – | – | – | – |
| **6** | **73.5** | $A_g$ | 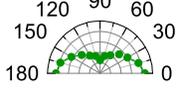 | 72.16 | $A_g$ | 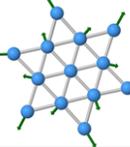 | 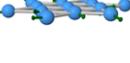 |
| – | – | – | – | 75.03 | $E_g$ | 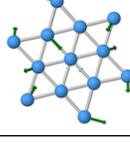 | 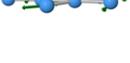 |
| 7 | 80.3 | $A_g$ | 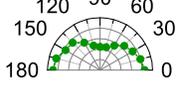 | 78.84 | $A_g$ | 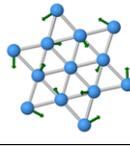 | 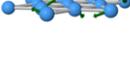 |
| 8 | 82.1 | $A_g$ | 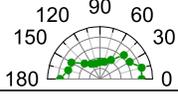 | – | – | – | – |
| **9** | **84.9** | $E_g$ | 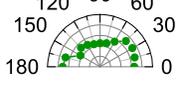 | 80.12 | $E_g$ | 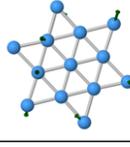 | 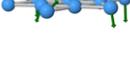 |
| – | – | – | – | 95.59 | $A_g$ | 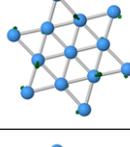 | 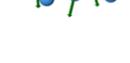 |
| **10** | **98.2** | $A_g$ | 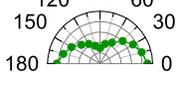 | 98.06 | $A_g$ | 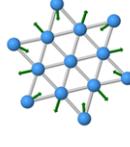 | 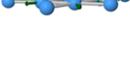 |

... table continued

| | Experiment | | | | Theory | | |
|---|---|---|---|---|---|---|---|
| Mode | $\omega$ (cm$^{-1}$) | Symm. | Polar | $\omega$ (cm$^{-1}$) | Symm. | Disp. $ab$ | Disp. $c$ |
| 11 | 101.9 | $A_\mathrm{g}$ | 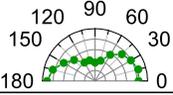 | – | – | – | – |
| – | – | – | – | 103.29 | $E_\mathrm{g}$ | 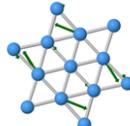 | 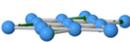 |
| 12 | 109.1 | $A_\mathrm{g}$ | 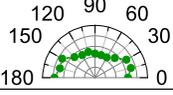 | – | – | – | – |

TABLE S2: **Assignment of the high frequency modes (140 - 300 cm$^{-1}$).** Selected modes of interest (no. 21) are highlighted in bold.

| | Experiment | | | | Theory | | |
|---|---|---|---|---|---|---|---|
| Mode | $\omega$ (cm$^{-1}$) | Symm. | Polar | $\omega$ (cm$^{-1}$) | Symm. | Disp. $ab$ | Disp. $c$ |
| – | – | – | – | 139.39 | $E_\mathrm{g}$ | 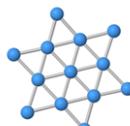 | 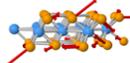 |
| – | – | – | – | 143.73 | $A_\mathrm{g}$ | 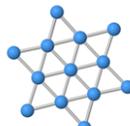 | 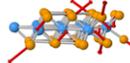 |
| – | – | – | – | 146.96 | $A_\mathrm{g}$ | 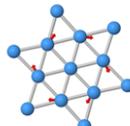 | 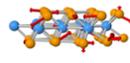 |
| 13 | 148.9 | $E_\mathrm{g}$ | 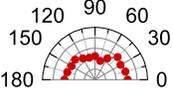 | 147.53 | $E_\mathrm{g}$ | 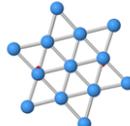 | 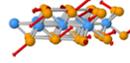 |
| 14 | 153.9 | $A_\mathrm{g}$ | 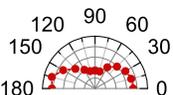 | 152.91 | $E_\mathrm{g}$ | 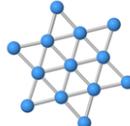 | 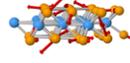 |

| **Experiment** | | | | **Theory** | | | |
|---|---|---|---|---|---|---|---|
| Mode | $\omega$ (cm$^{-1}$) | Symm. | Polar | $\omega$ (cm$^{-1}$) | Symm. | Disp. $ab$ | Disp. $c$ |
| – | – | – | – | 159.20 | $E_\mathrm{g}$ | 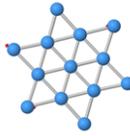 | 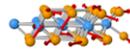 |
| 15 | 158.7 | $A_\mathrm{g}$ | 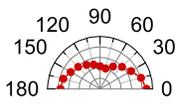 | 163.30 | $A_\mathrm{g}$ | 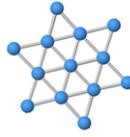 | 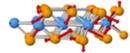 |
| 16 | 167.6 | $E_\mathrm{g}$ | 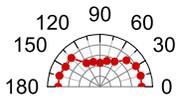 | 169.88 | $E_\mathrm{g}$ | 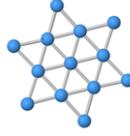 | 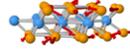 |
| – | – | – | – | 173.28 | $A_\mathrm{g}$ | 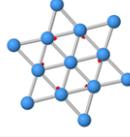 | 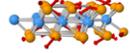 |
| 17 | 171.4 | $E_\mathrm{g}$ | 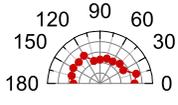 | 173.50 | $E_\mathrm{g}$ | 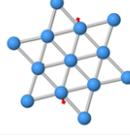 | 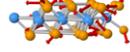 |
| 18 | 176.9 | $A_\mathrm{g}$ | 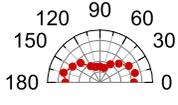 | 175.71 | $A_\mathrm{g}$ | 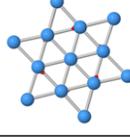 | 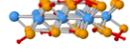 |
| – | – | – | – | 178.90 | $E_\mathrm{g}$ | | 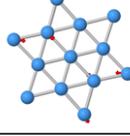 |
| 19 | 180.5 | $A_\mathrm{g}$ | 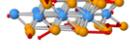 | – | – | – | – |
| – | – | – | – | 184.53 | $E_\mathrm{g}$ | 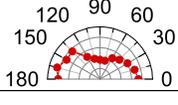 | 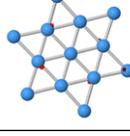 |
| 20 | 185.9 | $A_\mathrm{g}$ | 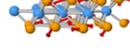 | 186.44 | $A_\mathrm{g}$ | 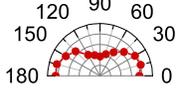 | 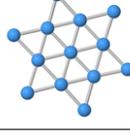 |

| Experiment | | | | Theory | | | |
|---|---|---|---|---|---|---|---|
| Mode | $\omega$ (cm$^{-1}$) | Symm. | Polar | $\omega$ (cm$^{-1}$) | Symm. | Disp. $ab$ | Disp. $c$ |
| **21** | **189.8** | $A_g$ | 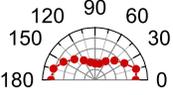 | 192.99 | $A_g$ | 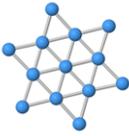 | 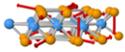 |
| 22 | 195.7 | $A_g$ | 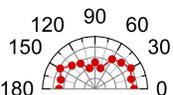 | 196.22 | $A_g$ | 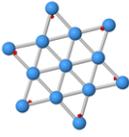 | 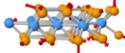 |
| – | – | – | – | 201.66 | $A_g$ | 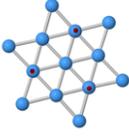 | 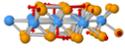 |
| 23 | 211.4 | $E_g$ | 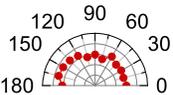 | 207.07 | $E_g$ | 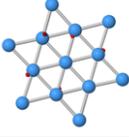 | 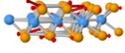 |
| – | – | – | – | 228.08 | $E_g$ | 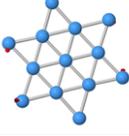 | 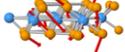 |
| – | – | – | – | 234.22 | $A_g$ | 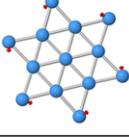 | 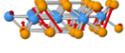 |
| 24 | 239.8 | $A_g$ | 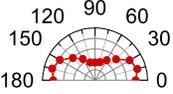 | 236.28 | $A_g$ | 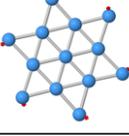 | 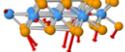 |
| – | – | – | – | 239.53 | $E_g$ | 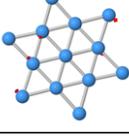 | 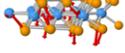 |
| – | – | – | – | 247.37 | $A_g$ | 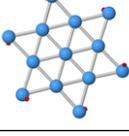 | 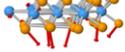 |
| 25 | 249.9 | $E_g$ | 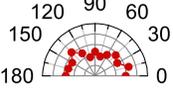 | 249.99 | $E_g$ | 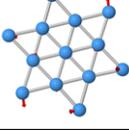 | 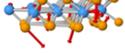 |

| Experiment | | | | Theory | | | |
|---|---|---|---|---|---|---|---|
| Mode | $\omega$ (cm$^{-1}$) | Symm. | Polar | $\omega$ (cm$^{-1}$) | Symm. | Disp. $ab$ | Disp. $c$ |
| 26 | 257.6 | $A_g$ | 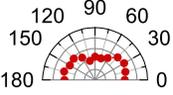 | 263.34 | $A_g$ | 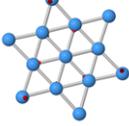 | 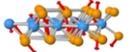 |
| 27 | 271.1 | $E_g$ | 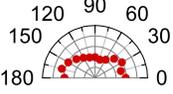 | 263.59 | $E_g$ | 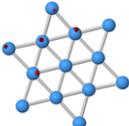 | 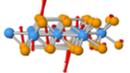 |

## C.  ANALYSIS OF ELECTRON-PHONON COUPLING

To assess the strength of electron-phonon coupling (EPC) in bulk $1T$-TaSe$_2$, we performed an analysis of the magnitude of perturbation of the electronic band structure for atomic displacements corresponding to the Raman-active modes, based on the results of the first-principles calculations discussed in Sections A & B. Displacements were created using the Phonopy MODULATION tag and band structures were calculated for the modulated structures in VASP, as described in the Methods section of the main text. Figure S3 shows the modulated structures induced by atomic displacements along the eigenvectors of the CCDW Γ-point phonons for the selected modes of interest (reported in bold font in Tables S1 and S2), where the colour scale represents the energy shift. The magnitude of the atomic displacement is arbitrary although here it results in range of absolute energy shift 0 to 7 meV across all modes in Figure S3. Here, we see that the 72.16 cm$^{-1}$ and 98.06 cm$^{-1}$ modes result in a particularly significant energy modulation of the bands across the entire Brillouin zone, indicating a strong EPC. In Figure S4, we compare the strength of EPC for modes in the low frequency region by plotting the absolute magnitude of the perturbations of the Fermi energy. The labelled phonon frequencies (55.40, 72.16, 78.48 and 98.06 cm$^{-1}$) correspond to modes with the strongest experimental Raman intensity. Clearly, we see a good agreement between the most intense Raman-active modes and the magnitude of the induced modulation of the Fermi energy. We emphasize that although these data are useful to highlight modes with strong EPC, they do not represent a complete prediction of

Raman intensities since the optical matrix elements for incident and emitted photons also enter into the Raman cross-section, which are not considered here.

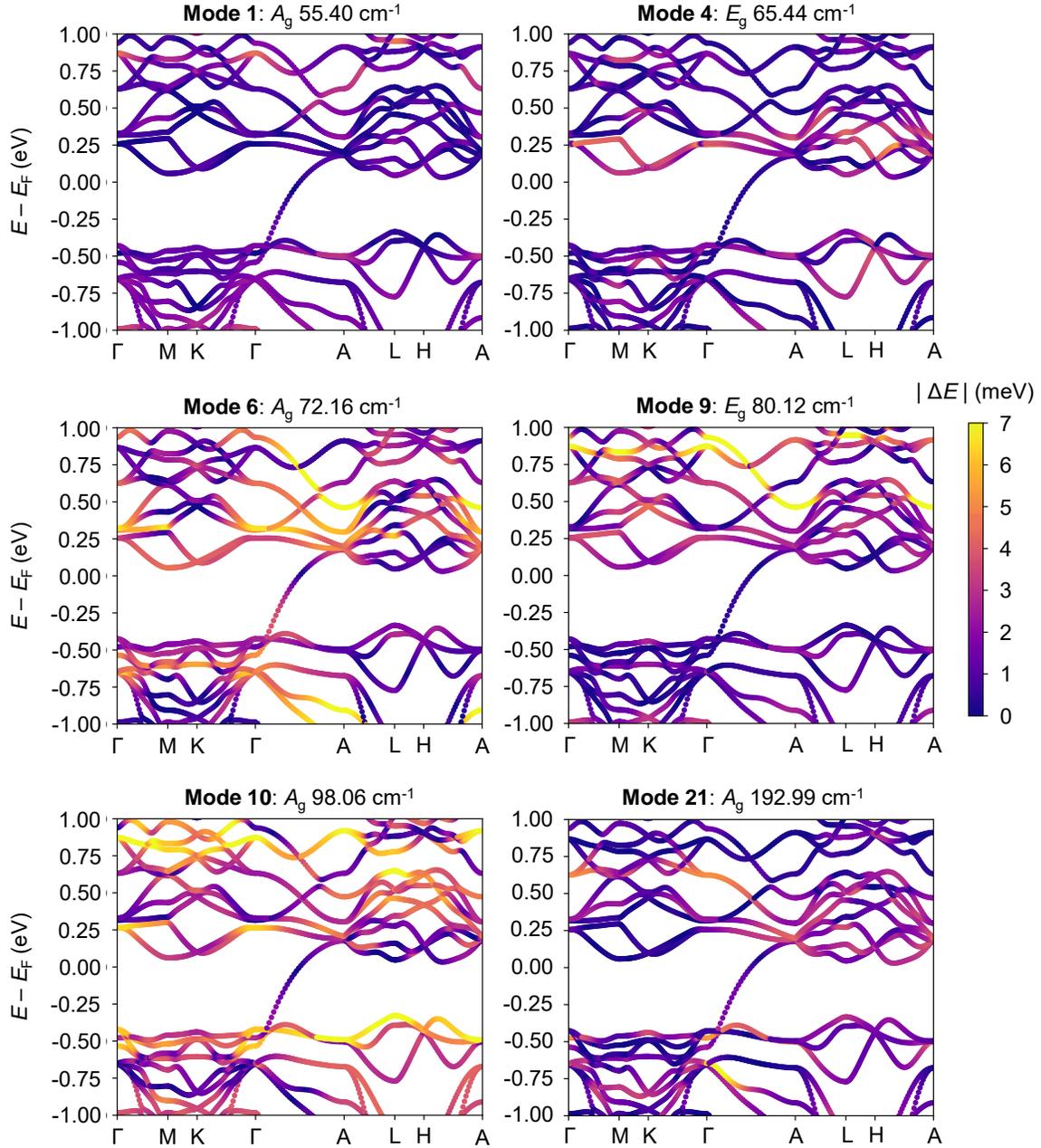

FIG. S3. **Phonon-modulated electronic band structures.** The colour scale represents the absolute energy modulation induced by atomic displacements along the eigenvectors of selected CCDW Γ-point phonon modes, as labelled.

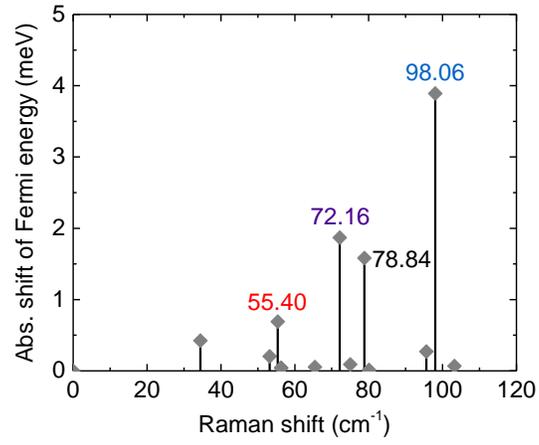

FIG. S4. **Comparison of EPC.** Absolute magnitude of the shift of the Fermi energy induced by phonons in the low frequency region. The labelled phonon frequencies (in cm$^{-1}$) correspond to modes with the strongest experimental Raman intensity.



\end{document}